\documentclass[onecolumn]{aastex62}
\usepackage{amsfonts, amsmath,amstext}
\usepackage{apjfonts}
\usepackage{natbib}
\usepackage{xspace}
\newcommand{\hnbsp}{\object{HN\,Peg\,B~}}
\newcommand{\hnb}{\object{HN\,Peg\,B}}
\newcommand{\editz}[1]{}
\newcommand{\teff}{\ensuremath{T_{\mathrm{eff}}}\xspace}

\begin{document}

\title{Cloud Atlas: Rotational Modulations in the L/T Transition Brown Dwarf Companion HN Peg B}

\shorttitle{HST Rotational Modulations of \hnb} \shortauthors{Zhou et al.}

\correspondingauthor{Yifan Zhou}
\email{yzhou@as.arizona.edu}

\author[0000-0003-2969-6040]{Yifan Zhou}
\altaffiliation{NASA Earth and Space Science Fellow}
\affiliation{Department of Astronomy/Steward Observatory, The University
  of Arizona, 933 N. Cherry Avenue, Tucson, AZ, 85721, USA}

\author{D\'aniel Apai} \affiliation{Department of Astronomy/Steward
  Observatory, The University of Arizona, 933 N. Cherry Avenue,
  Tucson, AZ, 85721, USA} \affiliation{Earths in Other Solar Systems
  Team, NASA Nexus for Exoplanet System Science.}
\affiliation{Department of Planetary Science/Lunar and Planetary
  Laboratory, The University of Arizona, 1640 E. University Boulevard,
  Tucson, AZ, 85718, USA}

\author{Stanimir Metchev}
\affiliation{Department of Physics \& Astronomy and Centre for Planetary Science and Exploration, The University of Western Ontario,
  London, Ontario N6A 3K7, Canada}
\affiliation{Department of Physics \& Astronomy, Stony Brook University, 100 Nicolls Rd, Stony Brook, NY 11794-3800, USA}

\author{Ben W. P. Lew}
\affiliation{Department of Astronomy/Steward Observatory, The University
  of Arizona, 933 N. Cherry Avenue, Tucson, AZ, 85721, USA}
\affiliation{Department of Planetary Science/Lunar and Planetary
  Laboratory, The University of Arizona, 1640 E. University Boulevard,
  Tucson, AZ, 85718, USA}

\author{Glenn Schneider}
\affiliation{Department of Astronomy/Steward Observatory, The University
  of Arizona, 933 N. Cherry Avenue, Tucson, AZ, 85721, USA}

\author{Mark S. Marley}
\affiliation{NASA Ames Research Center, Mail Stop 245-3, Moffett Field, CA 94035, USA}

\author{Theodora Karalidi}
\affiliation{Department of Astronomy and Astrophysics, University of California Santa Cruz,
  1156 High Street, Santa Cruz, CA 95064, USA}

\author{Elena Manjavacas}
\affiliation{Department of Astronomy/Steward Observatory, The University
  of Arizona, 933 N. Cherry Avenue, Tucson, AZ, 85721, USA}

\author{Luigi R. Bedin}
\affiliation{INAF - Osservatorio Astronomico di Padova, Vicolo
  dell'Osservatorio 5, I-35122 Padova, Italy}

\author{Nicolas B. Cowan}
\affiliation{Department of Earth \& Planetary Sciences and Department of Physics, McGill
  University, 3550 Rue University, Montr\'eal, Quebec H3A 0E8, Canada}

\author{Paulo A. Miles-P\'aez} \affiliation{Department of Physics \&
  Astronomy and Centre for Planetary Science and Exploration, The
  University of Western Ontario, London, Ontario N6A 3K7, Canada}
\affiliation{Department of Astronomy/Steward Observatory, The
  University of Arizona, 933 N. Cherry Avenue, Tucson, AZ, 85721, USA}

\author{Patrick J. Lowrance}
\affiliation{IPAC-Spitzer, MC 314-6, California Institute of
  Technology, Pasadena, CA 91125, USA}

\author{Jacqueline Radigan} \affiliation{Utah Valley University, 800
  West University Parkway, Orem, UT 84058, USA}

\author{Adam J. Burgasser} \affiliation{Center for Astrophysics and
  Space Science, University of California San Diego, La Jolla, CA
  92093, USA}

\begin{abstract}
  Time-resolved observations of brown dwarfs' rotational modulations provide powerful insights into the properties of condensate clouds in ultra-cool atmospheres. Multi-wavelength light curves reveal cloud vertical structures, condensate particle sizes, and cloud morphology, which directly constrain condensate cloud and atmospheric circulation models. We report results from Hubble Space Telescope/Wide Field Camera 3 near-infrared G141 taken in six consecutive orbits observations of \hnb, an L/T transition brown dwarf companion to a G0V type star. The best-fit sine wave to the $1.1-1.7\micron$ broadband light curve has the amplitude of $1.206\pm0.025\%$ and period of $15.4\pm0.5$ hr. The modulation amplitude has no detectable wavelength dependence except in the 1.4 \micron{} water absorption band, indicating that the characteristic condensate particle sizes are large ($>1\micron$). We detect significantly ($4.4\sigma$) lower modulation amplitude in the 1.4\micron{} water absorption band, and find that \hnb's spectral modulation resembles those of early T type brown dwarfs. We also describe a new empirical interpolation method to remove spectral contamination from the bright host star. This method may be applied in other high-contrast time-resolved observations with WFC3.
\end{abstract}


\section{Introduction}
Condensate clouds fundamentally impact the spectra, the pressure-temperature structure, the luminosity evolution, longitudinal and latitudinal temperature distribution, and energy transfer in the atmospheres of most transiting exoplanets \citep[e.g.,][]{Kreidberg2014,Sing2016,Stevenson2016}, directly imaged exoplanets \citep[e.g.,][]{Skemer2014, Ingraham2014, Bonnefoy2016} and brown dwarfs \citep[e.g.,][]{Marley2002, Burgasser2002, Knapp2004}. Therefore, parametrized cloud models are an essential but not well tested component of atmospheric models. They play a particularly important role in many atmospheric retrieval studies \citep[e.g.][]{Benneke2012,Line2012,Line2015,Burningham2017}.  However, cloud properties are highly degenerate in disk-integrated observations with other global and difficult-to-measure parameters (e.g., bulk composition, vertical mixing rate, surface gravity, non-equilibrium chemistry). Time-resolved observations of rotational modulations enable disentangling the effects of global parameters (constant for a given object) from locally varying parameters (primarily cloud coverage), thus providing a powerful method for testing cloud models. Near-infrared photometric variability samples longitudinal cloud cover distribution \citep[e.g.,][]{Artigau2009, Radigan2012, Metchev2015, Zhou2016}, and high-dispersion spectroscopy can provide two-dimensional maps \citep{Crossfield2014} and sample opacity variations in single atomic or molecular species \citep[e.g.,][]{Kellogg2017}. Time-resolved low- and medium dispersion space-borne spectroscopy - with high cadence and signal-to-noise ratio - can locate the pressure levels of the condensate clouds and reveal cloud vertical structures \citep[e.g.,][]{Apai2013,Yang2014,Yang2016, Schlawin2017}.
An important prediction of most cloud models is that surface gravity -- through its impacts on pressure scale height and dust settling rate -- will have profound effects on cloud thickness. Indeed, exceptionally thick clouds have been proposed as the origin of the very red colors and low near-infrared luminosity of several directly imaged exoplanets \citep[e.g.,][]{Skemer2011,Marley2012}.

Recent results further increased the potential of time-resolved observations and rotational mapping for brown dwarf and exoplanet atmospheric characterization. \citet[][]{Karalidi2015} developed \textit{Aeolus}, which retrieves two-dimensional top-of-the-atmosphere maps from disk-integrated light curves. \citet[][]{Apai2017} identified bands and spots in brown dwarf atmospheres and demonstrated similarities between atmospheric circulations in L/T transition brown dwarfs and in Neptune. The discovery of rotational modulations in directly imaged exoplanets/planetary-mass objects \citep{Biller2015, Zhou2016} allows comparative studies of condensate clouds in brown dwarfs and exoplanets.


\textit{Cloud Atlas} is a Hubble Space Telescope (HST) Wide field Camera 3 (WFC3) Large Treasury program (Program No.: 14241; PI: Apai). The primary goal of the Cloud Atlas project is to identify the role of surface gravity in setting the properties of condensate clouds. To achieve this goal, the project selected 21 brown dwarfs and planetary-mass companions that were divided into four groups, (i) high effective temperature (\teff) and high surface gravity ($g$), (ii) high \teff and low $g$, (iii) low $\teff$ and high $g$, and (iv) low $\teff$ and low $g$. We scheduled time-resolved spectroscopic observations for isolated or low-contrast targets and photometric observations for high-contrast targets.  Each object was initially observed in two consecutive HST orbits to assess the presence of amplitude variability, then down-selecting a sub-set of objects to study with deep-look observations (DLO) with 6--12 follow-on consecutive orbits. We paid particular attention to the difference in rotational modulations in and out of the 1.4\,\micron{} water absorption band because it is a sensible probe to cloud vertical structure \citep{Yang2014}.  HN Peg B was among the targets we selected for six consecutive orbits DLO. Details of Cloud Atlas observations can be found on \url{http://www.stsci.edu/cgi-bin/get-proposal-info?observatory=HST\&id=14241} and on \url{http://apai.space/cloudatlas}

\hnbsp \citep{Luhman2007} is T2.5 type brown dwarf companion to its G0V type host star. \hnbsp has a projected angular separation of $43.2''\pm0.4$ from its host star, which corresponds to project physical distance of $795\pm15$ au \citep{Luhman2007}. \hnbsp and HN Peg A have a brightness contrast of 11.07 magnitude in $J$ band. The mass of \hnbsp is estimated to be 12 -- 30 $M_\mathrm{Jup}$ based on evolutionary tracks \citep{Luhman2007, Leggett2008}. Upon its discovery, \citet{Luhman2007} classified HN Peg B to have low or intermediate surface gravity based on the youth of the host star (e.g., \citealt{Gaidos1998}, 200-800\,Myr; \citealt{Barnes2007}, $237\pm33$\,Myr). However, \citet{Leggett2008} found that the near infrared spectra of HN Peg B agree better with higher gravity template ($\log g=4.81$), which casts doubt on the original low surface gravity classification. Nevertheless, we tentatively included \hnbsp in the low~$g$ low \teff{} category because it belongs to a small sample of T dwarf that show at least partial evidence of low surface gravity.

Using \textit{Spitzer Space Telescope} time-resolved photometry, \citet{Metchev2015} discovered rotational modulations in both the [3.6] and the [4.5] channels, broadband light curves of \hnb. \citet{Metchev2015} classified the variability period type as \emph{long} and used a Fourier series fit to determine the rotation period to be $18\pm4$ hr, close to the total observation length. The discovery of its rotational modulations made \hnbsp an ideal brown dwarf companion for time-resolved spectroscopy.

\section{Observations and Data Reduction} We observed \hnbsp using \textit{HST}/Wide Field Camera 3 near infrared (\textit{HST}/WFC3/IR) channel on 2017 May 16 as part of the HST Treasury program {\em Cloud Atlas}. We monitored the target using the G141 low-resolution ($R\sim130$ at $1.4 \micron$) grism in six consecutive orbits (8.6 hrs time baseline). Each orbit contained 10 consecutive G141 spectroscopic frames and one direct image frame using the F132N filter to aid wavelength calibrations. All exposures were taken with the $256\times256$ sub-array to reduce on-board memory requirements to avoid in-orbit buffer downloads. The exposure time for the spectroscopic frames was 201.40 s, resulting in a cadence of 220 seconds. We did not apply dithering to our observations to avoid systematics arising from flat-field uncertainties that could limit photometric precision. The positions of the spectral images show sub-pixel shifts due to HST's pointing drift. By cross-correlating each frame to the first one, we found that the largest shifts were 0.02 pixel along the $x$-axis and 0.03 pixel along the $y$-axis. These shifts were much smaller than the 8-pixel extraction window, so the potential systematics caused by the drift are negligible.

In order to minimize the contamination to \hnb's observed spectrum by its host star, we constrained the roll angle of the telescope to separate the primary and companion spectra on the detector.  However, instrumentally scattered grism-dispersed light from the bright host star is also distributed in a complex band-like pattern across the field-of-view that contaminates the traditional sky background at the location of the companion spectrum (see Figure~\ref{fig:band}). At the location of the maximum, HN Peg A's contamination pattern contributes about 25\% of the pixel counts and, without mitigation, degrades the precision of the spectral and variability measurements. In \S~\ref{sec:removal}, we describe in detail how we removed the contamination band pattern.
 
We started our reduction with the CalWFC3 product \texttt{flt} frames. Most data reduction procedures were done using an aXe-based \citep{Kummel2009} pipeline following previous HST/WFC3 brown dwarf time-resolved spectral observation studies \citep[e.g.,][]{Apai2013,Lew2016}. However, after the sky background subtraction, we included an additional step (described below) to remove the band pattern. After this step, we followed the regular reduction approach, i.e., we fed the band-subtracted frames to aXe and extracted the spectral sequence.

The extracted light curve for \hnbsp showed easily recognizable signatures of ramp effect systematics. These systematics were widely reported in time-resolved HST/WFC3 observations\citep[e.g., ][]{Berta2012, Apai2013, Deming2013}. We successfully removed these systematics in the band-subtracted light curves using the solid state physics-motivated RECTE charge trap correction method \citep{Zhou2017}.

\subsection{Primary Star Contamination Removal: Procedures}
\label{sec:removal}
\begin{figure*}[th]
  \centering
  \plottwo{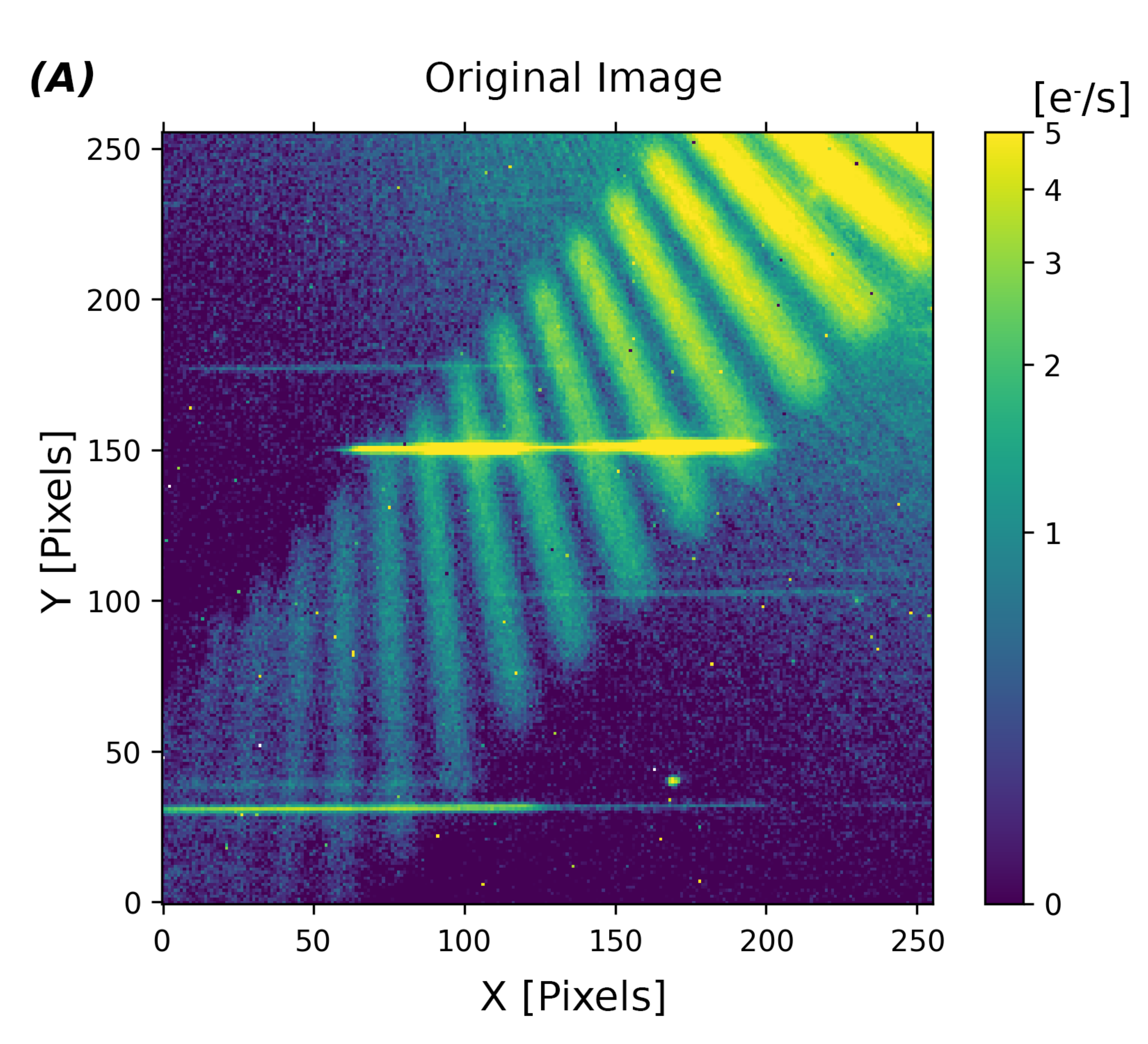}{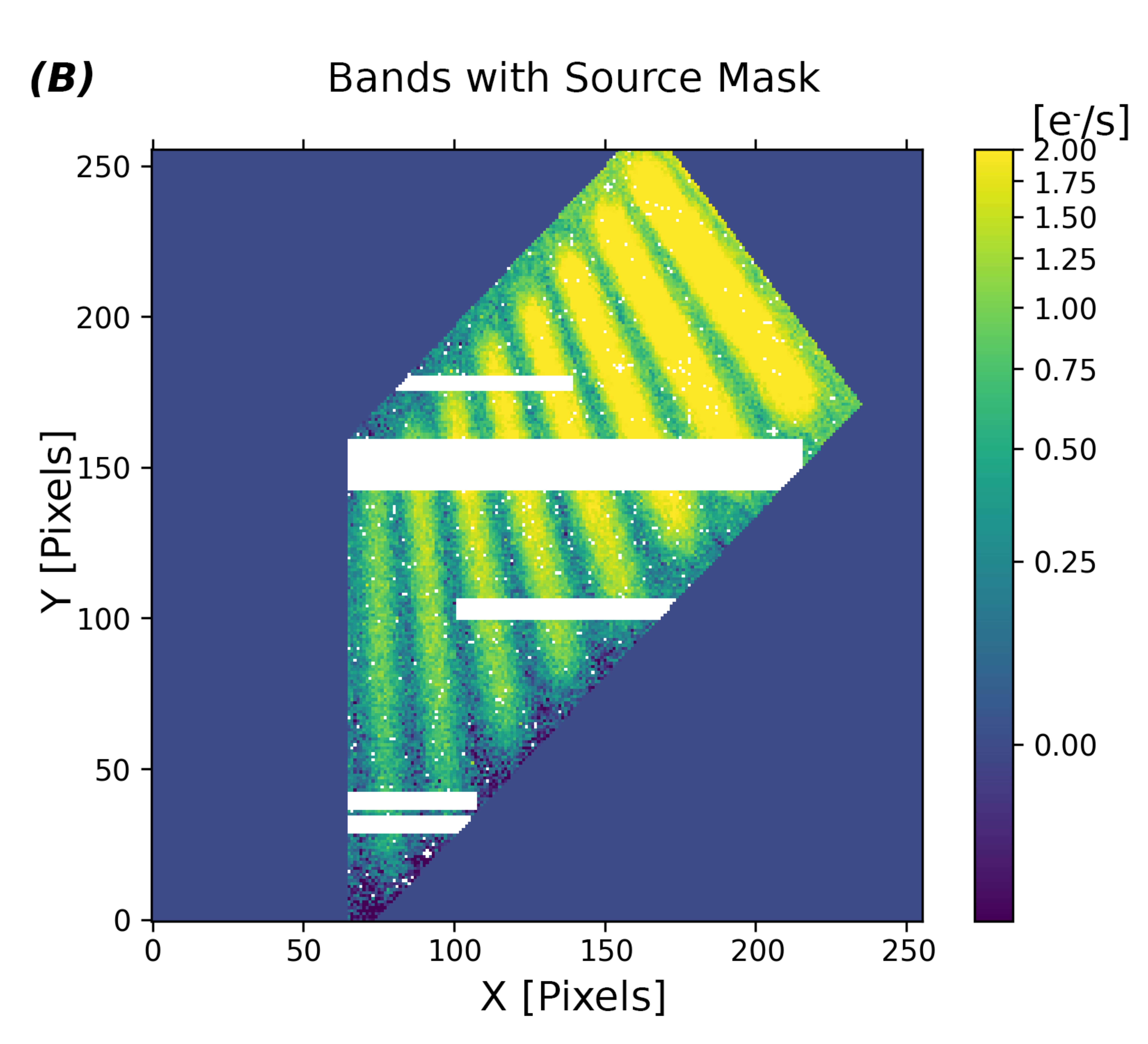}
  \plottwo{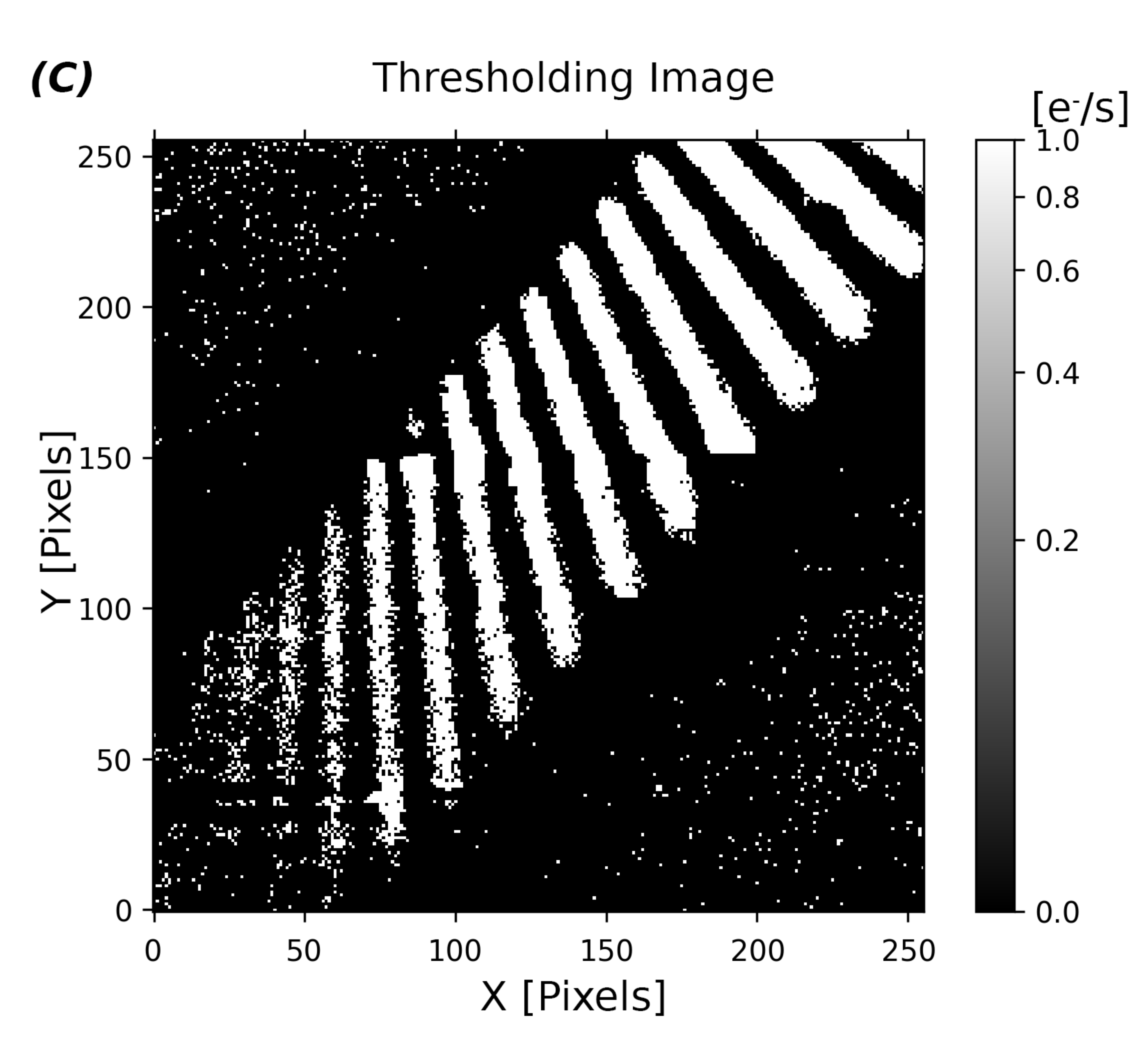}{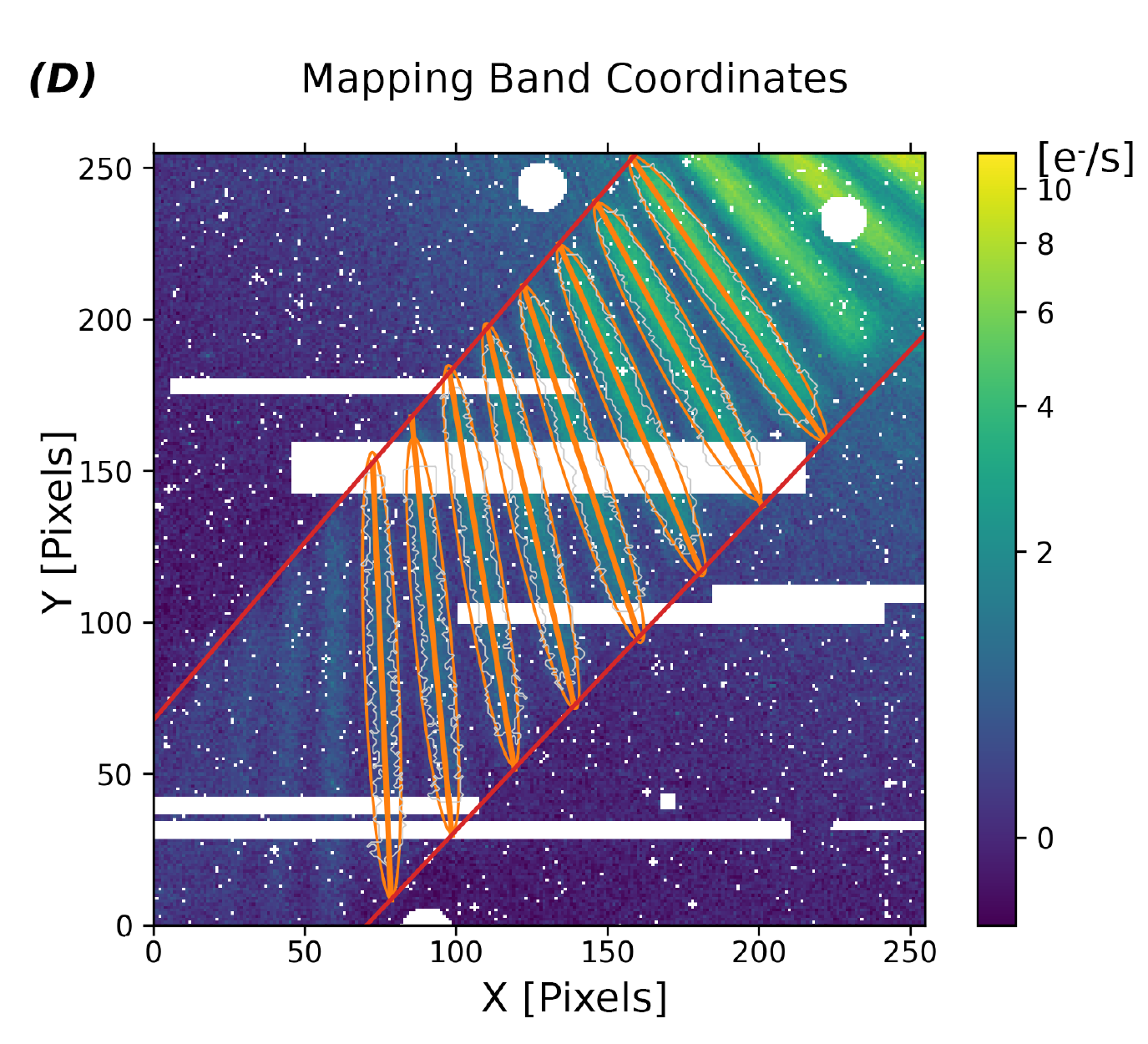}
  \plottwo{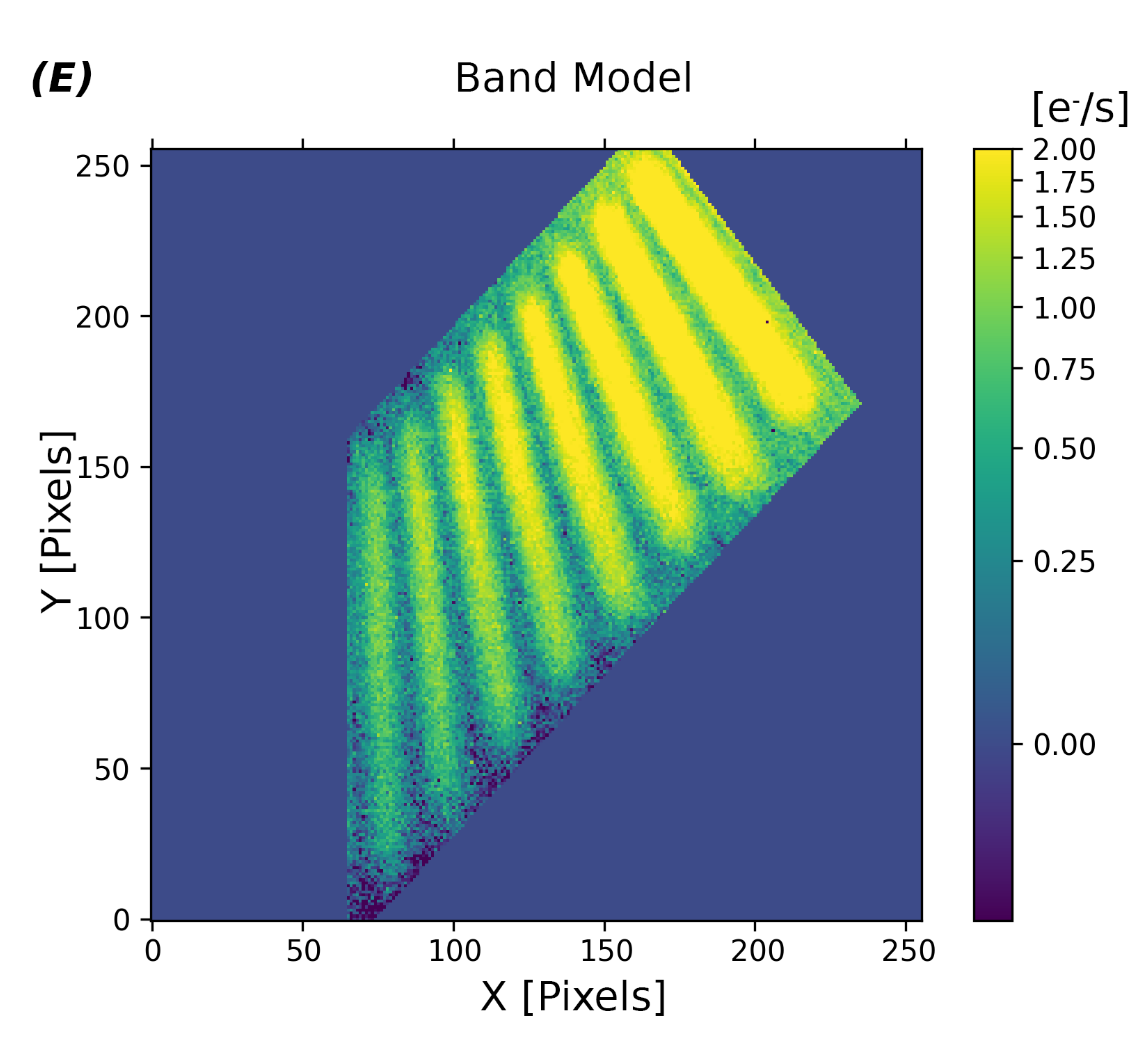}{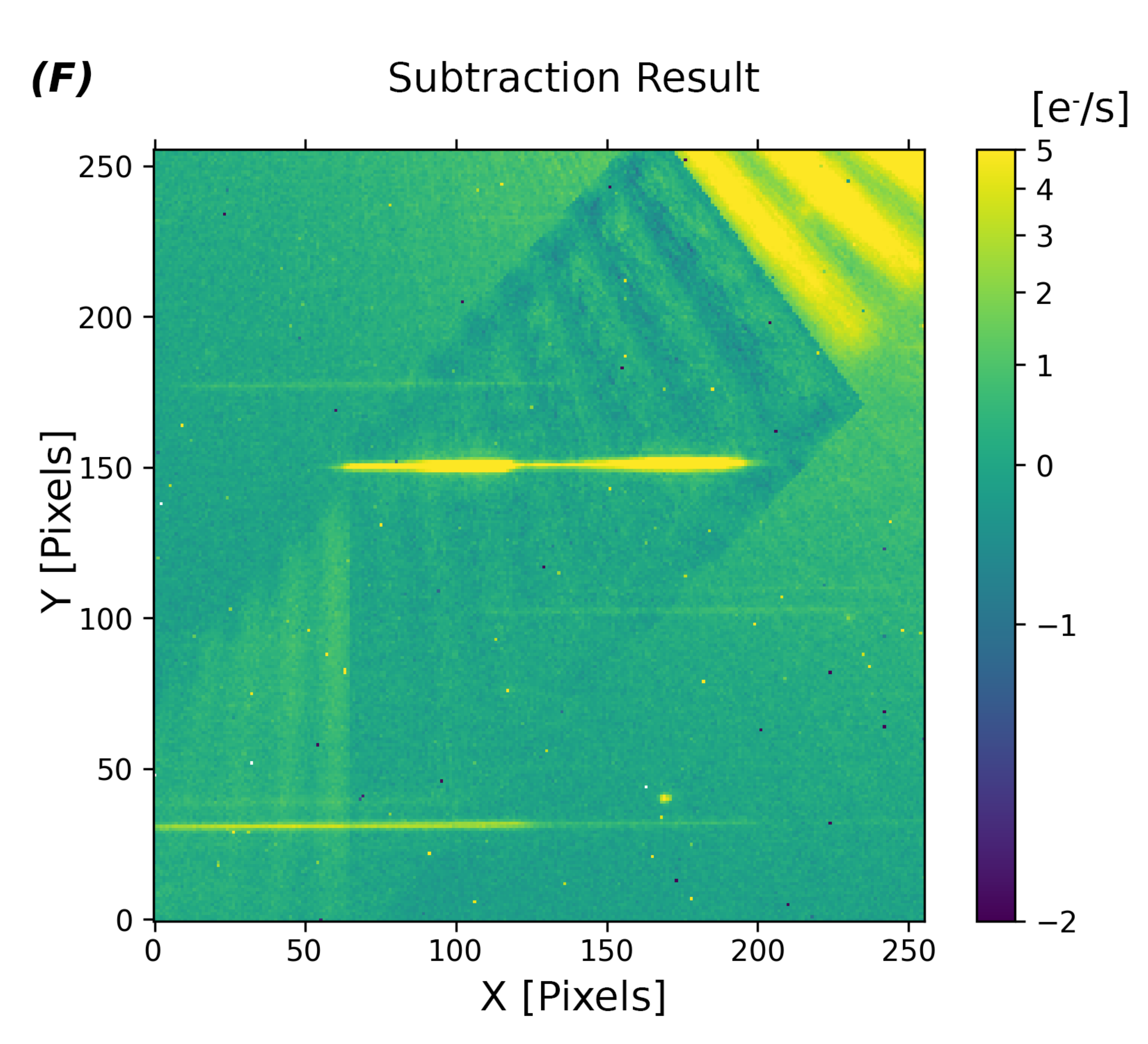}
  \caption{\emph{An example for original \texttt{flt} frame,
      intermediate product, and the result after band subtraction.}
    The order of the images corresponds to the sequence of reduction
    steps. The Images are: A) image before band subtraction; B) high
    S/N bands with astrophysical signal masked out; C) thresholding
    results, foreground pixels are plotted in white; D) coordinate
    regularization and re-mapping; E) empirically interpolated
    surface; F) image after band subtraction. }
  \label{fig:band}
\end{figure*}

\begin{figure*}[!t]
  \centering
  \includegraphics[height=0.36\textwidth]{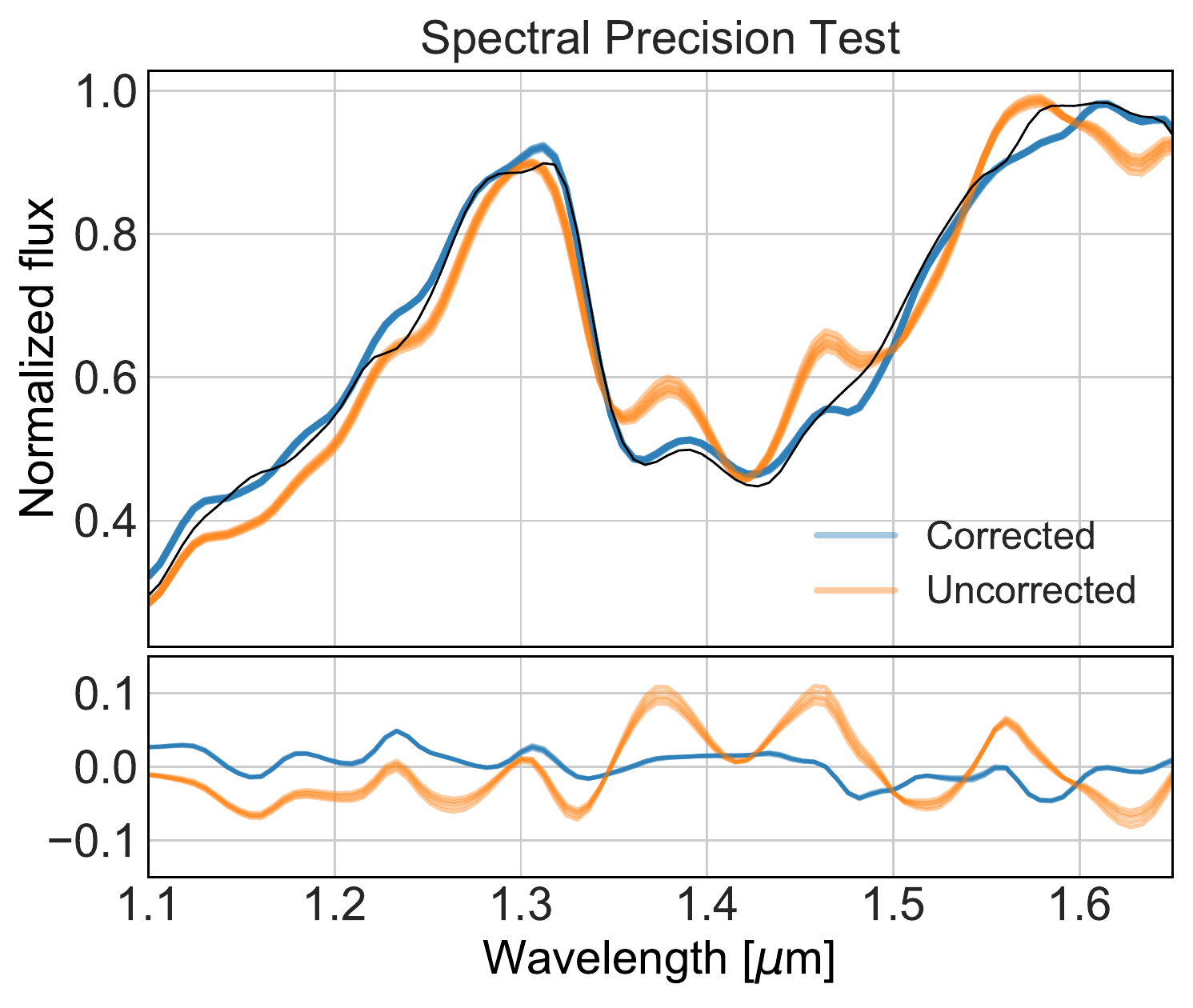}  
  \includegraphics[height=0.36\textwidth]{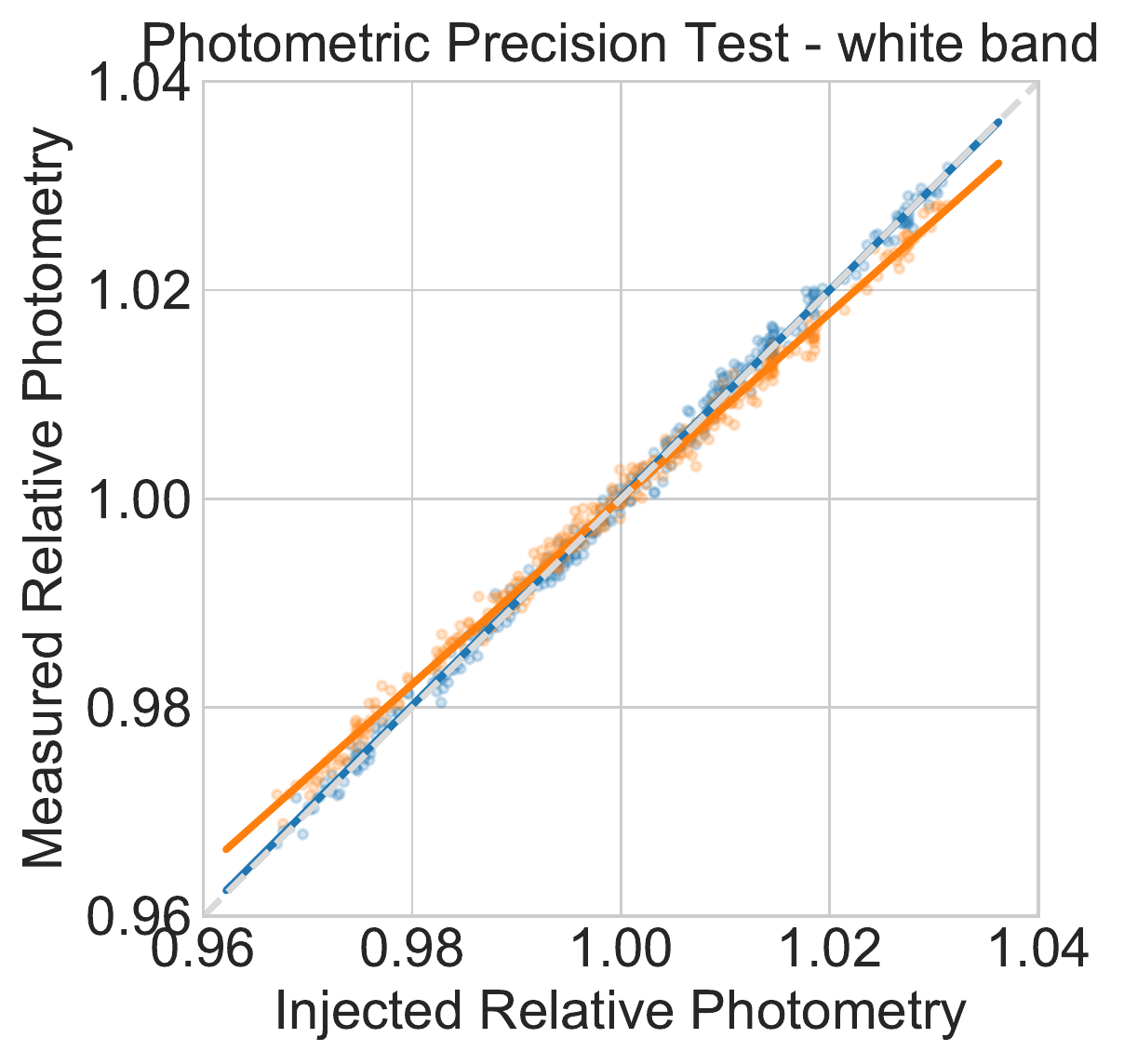}\\
  \includegraphics[width=0.32\textwidth]{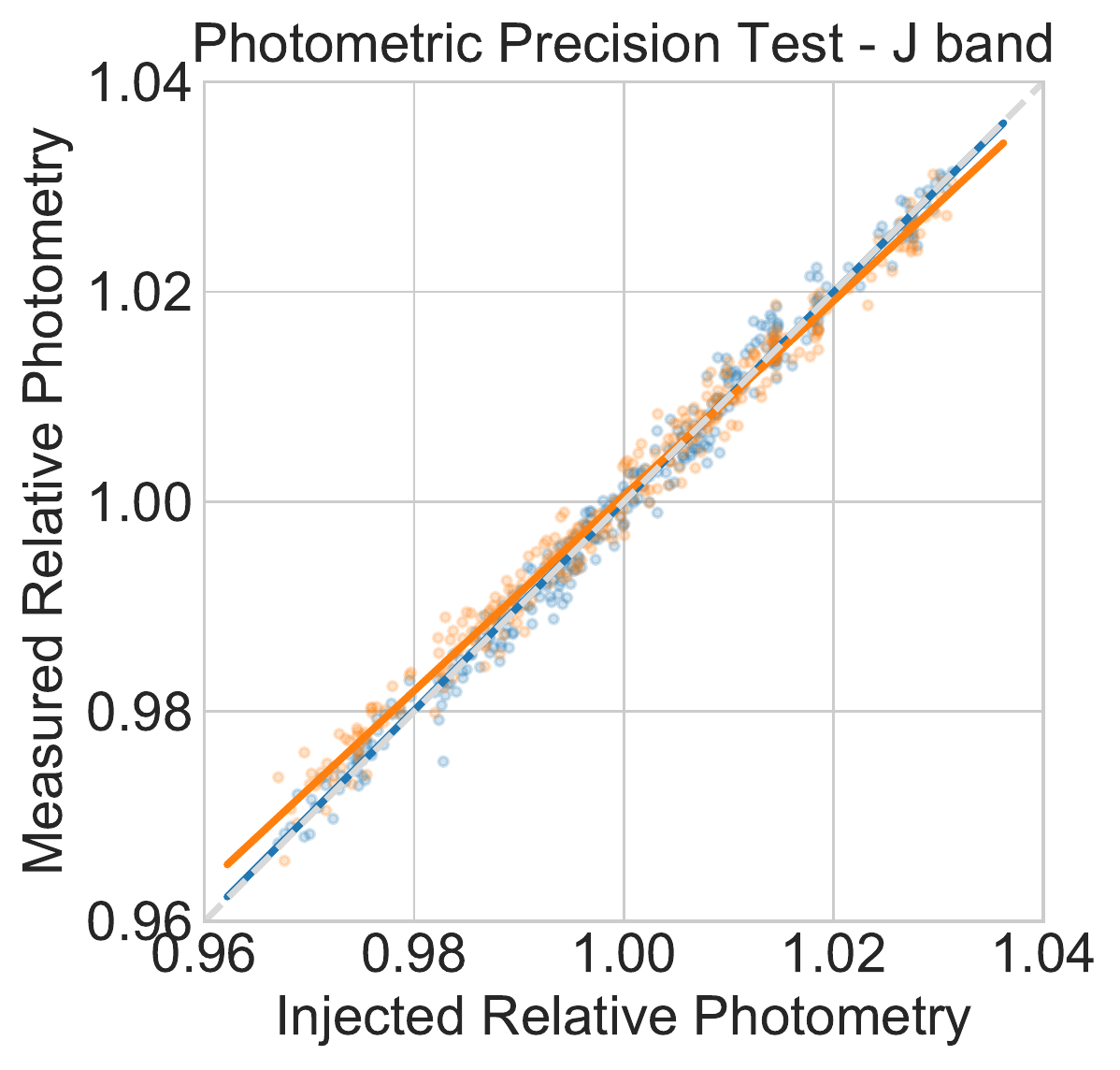}
  \includegraphics[width=0.32\textwidth]{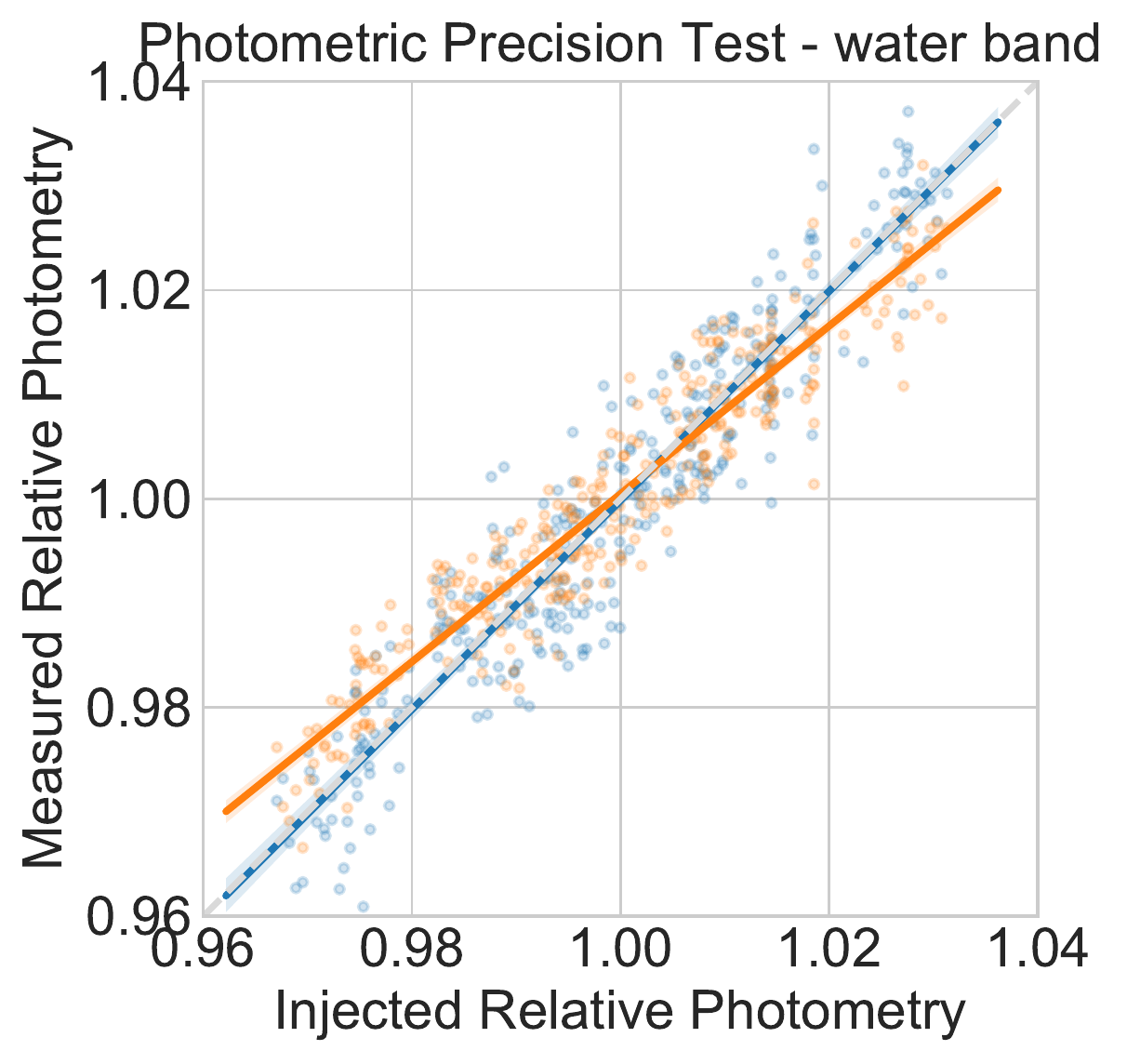}
  \includegraphics[width=0.32\textwidth]{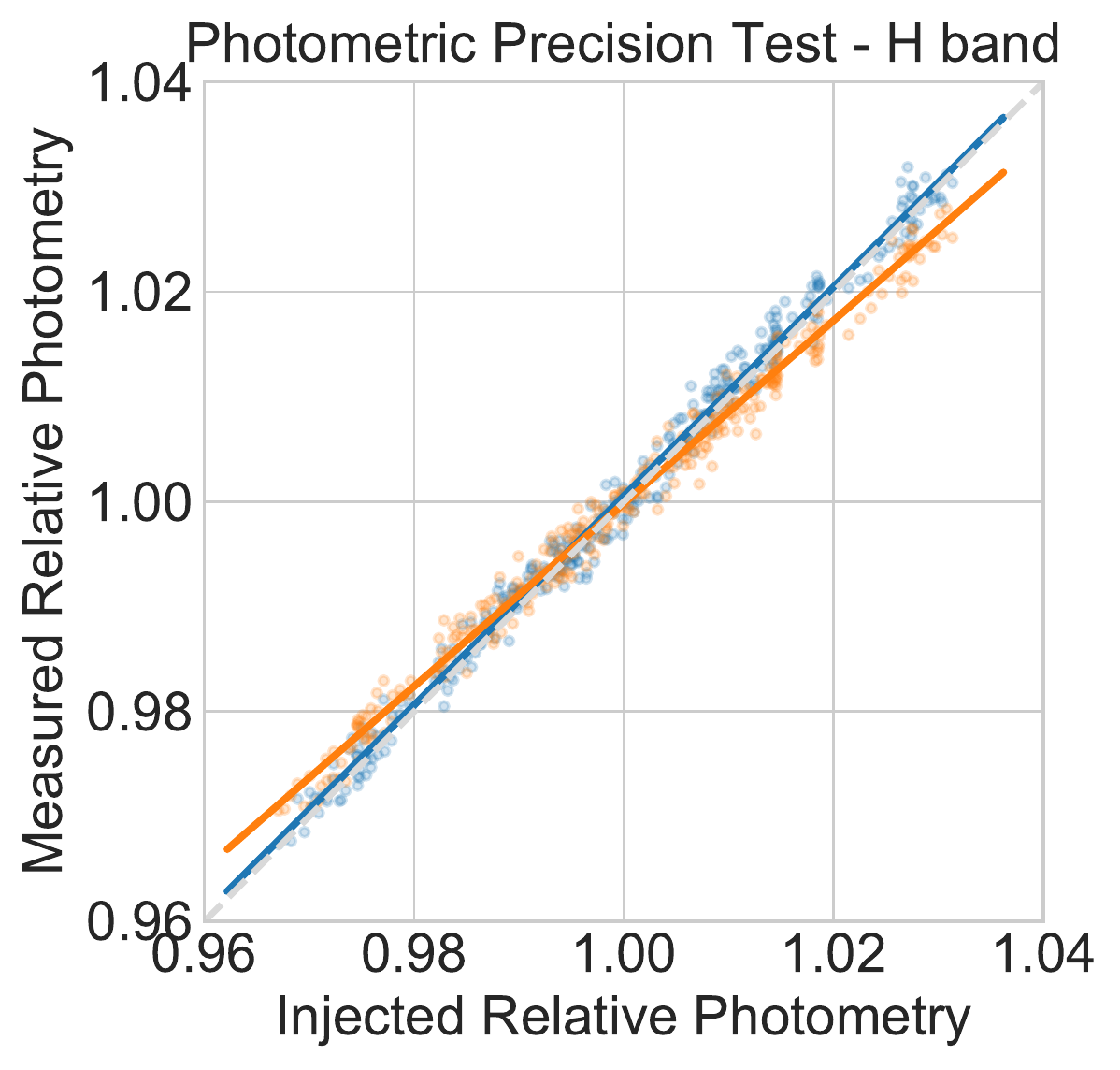}
  \caption{\emph{The effect of band subtraction}. \emph{Upper left}:
    comparison of spectral recovery. Blue curves are extracted spectra
    after band subtraction and orange curves are those without band
    subtraction. Injected spectra are plotted in black
    lines. \emph{Upper right}: comparison of broad-band photometry
    recovery. Extracted relative broadband photometry is plotted
    against the injected signal. Perfect recoveries (1:1) are plotted
    in gray dashed lines for references. \emph{Lower}: The same plots as
    in the upper left panel, but for the $J$, 1.4\micron-water, and $H$ bands.}
  \label{fig:eval}
\end{figure*}

Thanks to the well-defined and repeatable spatial variations of the
contamination pattern, we successfully removed them using an
empirical interpolation method.

Our band subtraction algorithm includes three steps: (1) band recognition, (2) mapping the bands to a coordinate system that is established based on band structure gradient, and (3) interpolation. Image processing tasks that we used are available in python package \textit{scikit-image} \citep[][]{scikit-image}. First, the algorithm recognizes and isolates the bands from the rest of the image structure. To avoid the confusion of bands with the spectrum of \hnbsp we pre-processed every frame by conservatively masking the spectral trace's of \hnbsp and background stars (Figure~\ref{fig:band} panel B). We then use ''inpainting" algorithms \citep{Bertalmio2001} to interpolate masked image regions. ``inpainting'' algorithms reconstruct masked pixels based on non-masked region and improve the precision in recognizing the bands.  We then segmented the bands and the background using locally optimized thresholding method. The algorithm computed a threshold mask based on local pixel neighborhood, which effectively marked the foreground pixels (band pattern) as ``1'' and background pixels as ``0'' (Figure~\ref{fig:band} panel C). Contours on the binary images were then used to identify individual bands. To filter out low signal-to-noise ratio (SNR) detection, we selected contours that were entirely included in the image and had enclosed sizes that are larger than 30 percentile of those of all detected contours. This procedure resulted in eight bands identified per frame (Figure\,\ref{fig:band} panel $D$).

Second, we regularized the selected bands and mapped them with a unified coordinate system. For each band, we started the regularization by identifying the ``semi-major axis'', defined as the semi-major axis of an ellipse with the same normalized second central moment as the band region.  The ``semi-major axis'' measurement is a good estimate of the length and orientation of the bands \citep{Chaumette2004}. We assumed that the top and bottom endpoints of each ``semi-major axis'' lay on two lines. Therefore, we performed linear regressions to the top and bottom axis endpoints. The endpoints were then adjusted along the axes such that the respective points joined contiguously and co-linearly without any discontinuity. We then established an orthogonal coordinate system for each band. We converted the $(x, y)$ image coordinates of each pixel to coordinates $(\rho, r)$ in which $\rho$ is the distance from the upper end of the band in the ``semi-major axis'' direction and $r$ is the distance from the point to the semi-major-axis. In addition, we normalized $\rho$ to the length of the semi-major-axis to account for the individual length of each band. The new coordinate systems were established for individual bands, and the axes of the coordinate systems were aligned with gradients of the surface brightness of each band.

Third, we used empirically determined interpolation functions to calculate the pixel values where bands and astrophysical spectra overlapped; i.e. in the regions masked as shown in Figure~\ref{fig:band} panel B. We fit a bi-cubic spline surface to each band.  Numbers of knots for the cubic splines were 5 and 4 for $r$ and $\rho$ direction, respectively.  Finally, the best-fitting spline surface was the model for band intensity distribution (Figure~\ref{fig:band} panel E) and was used for its subtraction (Figure~\ref{fig:band} panel F).

\subsection{Contamination Removal: Error Analysis}

We evaluated the band subtraction quality by injecting synthetic spectra to the original data and then measuring the difference of the injected and extracted signals with and without band subtraction.  Using the software aXeSIM \citep{Kummel2009} we generated four sets of synthetic spectra that shared similar brightness and spectral energy distributions as \hnb. The synthetic spectra had $y$-axis displacements of $\Delta y=+40,\,+15,\,-20,\,-30$ pixels. We injected simulated rotational modulation signals by multiplying the synthetic spectra with a sinusoidal time series and adding the products to the original frames. When adding the synthetic spectra, we included random noise that had the same standard deviations as the photon noise. We then processed the synthetic datasets in two ways, once with band removal and the once without it. We performed ten iterations in a Monte Carlo fashion. For each iteration, the injected sinusoidal signal had random amplitude, period, and phase. The results are presented in Figure~\ref{fig:eval}.

Based on these tests, we conclude that the band subtraction improved the precision of the spectra and of the light curves. As shown in Figure~\ref{fig:eval}, without band subtraction, the background structures skewed the spectral measurements by up to 10\% of the injected spectra. After band subtraction, the difference between the injected and the measured spectra were on the sub-percent levels. For the light curve measurements, we defined the correction improvement ($I$) as the relative reduction of root mean square residuals between measured photometry and injected photometry (Equation~\ref{eq:3}).
\begin{equation}
  \label{eq:3}
  I = 1-\frac{RMS(\mathrm{Phot_{corr}}-\mathrm{Phot_{inject}})}
    {RMS(\mathrm{Phot_{uncorr}}-\mathrm{Phot_{inject}})}
\end{equation}
in which $\mathrm{Phot_{inject}}$, $\mathrm{Phot_{uncorr}}$, and $\mathrm{Phot_{corr}}$ are injected photometry, photometry measured in uncorrected frames and photometry measured in corrected frames. We calculated $I$ for four photometric bands: $1.1-1.7$ white, $J$, $H$, and 1.4 water. For the four synthetic data sets $I$ values for white band were similar and ranged from 45\%$-$55\%. $I$ values for the $J$, $H$, and water bands varied because of the different $y$ positions of the spectra resulted in different overlapping spectral regions. Nevertheless, none of the bands did the root mean square residuals between measured photometry and injected photometry exceed photon noise by 10\% after correction. Because background structures introduced a constant flux to each pixel, the measured relative modulation amplitudes were reduced by as much as 10\%. After band subtraction, the measurements were able to recover the true injected signal. These tests verified that our band subtraction for the spectral modulation measurement was highly precise and demonstrated the importance of careful background analysis and systematics removal for high-precision time-resolved observations.

\section{Keck/NIRSPEC Observations of HN Peg B}
\begin{figure}[!t]
  \centering
  \includegraphics[width=\columnwidth]{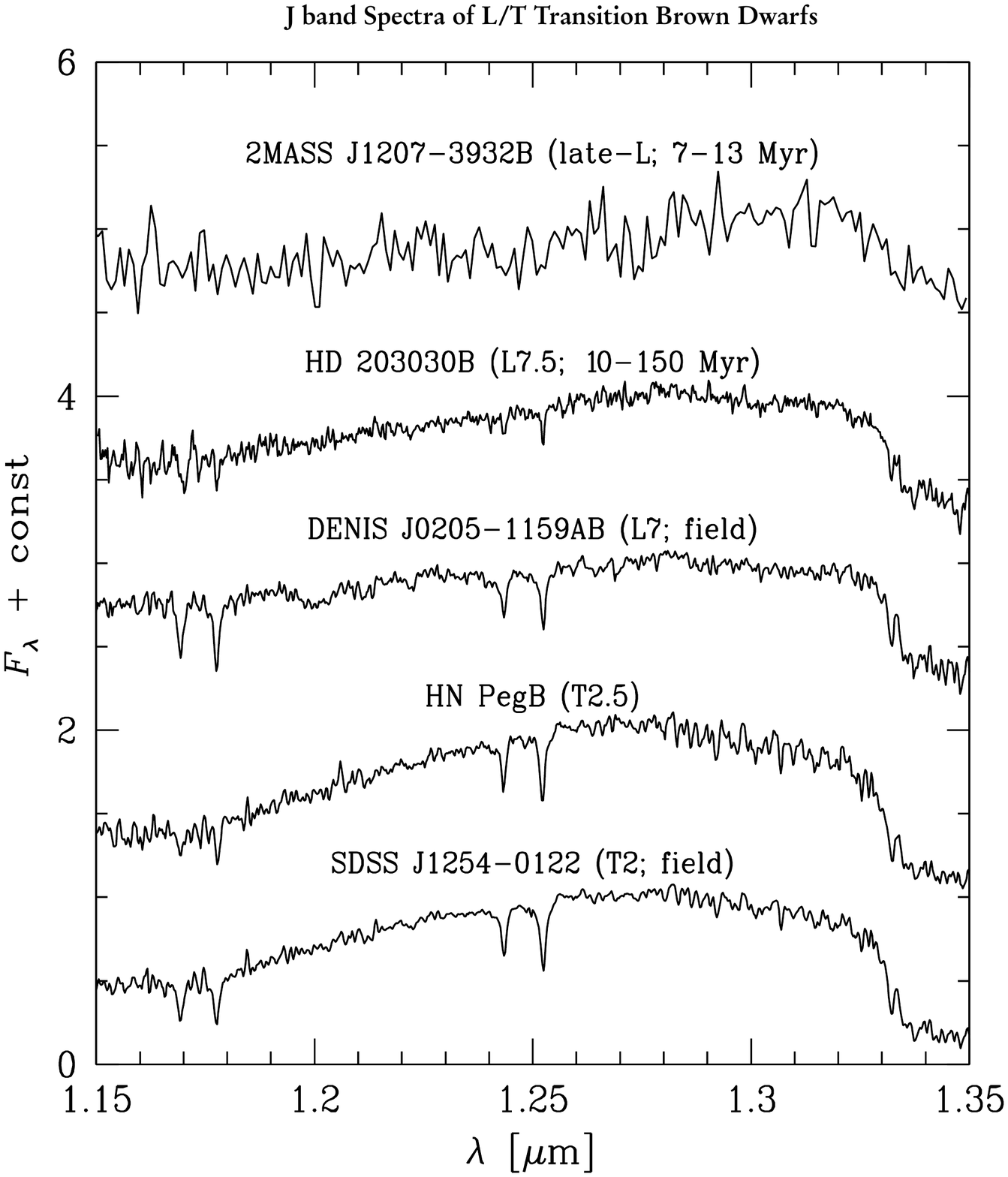}
  \caption{The Keck/NIRSPEC $J$-band spectrum of HN\,Peg\,B compared to
    the spectra of young and field-aged L/T transition dwarfs: the
    10$\pm$3 Myr-old late-L TWA member 2MASS J1207$-$3932B
    \citep{Chauvin2004}, the 10-150 Myr-old L7.5 dwarf HD 203030B
    \citep{Metchev2006,Miles2017}, and the field-age ($>$500 Myr old)
    L7 and T2 dwarfs DENIS J0205–1159AB and SDSS J1254–0122 (spectra
    from \citealt{McLean2003}). $J$ band spectra for L/T transition brown
    dwarfs. HN\,Peg\,B resembles high surface gravity field brown
    dwarfs.}
  \label{fig:spec}
\end{figure}

We also present here a moderate-resolution ($R\approx2300$) $J$-band spectrum of \hnbsp that was obtained with Keck/NIRSPEC on 2008 July 8.  We used the N3 (1.143--1.375~$\micron$) filter with a two-pixel ($0\farcs38$) wide slit, and exposed for a single ABBA sequence totalling 40 minutes of integration time.  Standard stars and arc lamps were observed after the target. We performed the preliminary data reduction with the REDSPEC pipeline \citep[REDSPEC Data Reduction Manual\footnote{ \href{http://www2.keck.hawaii.edu/inst/nirspec/redspec.html}{http://www2.keck.hawaii.edu/inst/nirspec/redspec.html}}]{Prato2002}.  Individual exposures were flat-fielded, rectified, and wavelength calibrated. Optimal extraction of the spectra was done with the IRAF {\sc apall} package. After correcting for telluric absorption the individual spectra were median combined.  The final NIRSPEC N3 spectrum of \hnbsp is shown in Figure~\ref{fig:spec}, where it is compared to the spectra of 100--150\,Myr-old late-L dwarfs and of field ($>500$\,Myr-old) L7 and T2 dwarfs.

As shown in Figure~\ref{fig:spec}, the two \ion{K}{1} absorption doublets at $\lambda \simeq 1.1692$-$1.1778\,\micron$ and $\lambda \simeq 1.2437$-$1.2529\,\micron$ are well-established surface gravity-sensitive diagnostics for distinguishing young ($\lesssim150$ Myr-old) from field-aged ($\gtrsim$500~Myr) ultra-cool dwarfs \citep{Allers2013,Liu2016}.  The comparison among the $J$-band spectra (Figure~\ref{fig:spec}) shows these lines in HN\,Peg\,B to be comparable in strength to the field-aged objects and much stronger (more pressure-broadened) than in the young objects.  We therefore conclude that the age of HN\,Peg\,B is likely $\gtrsim$500 Myr.  This is consistent with the findings of \citet{Luhman2007}, who assign a 100-500 Myr age for the primary HN\,Peg\,A from chromospheric activity and space kinematics arguments.  The age also agrees well with the conclusions of \citep{Leggett2008}, wherein they find best-fit to 1--4 micron spectral models (e.g., their Fig 4) with moderately high surface gravities ($\log\,g=4.8$) in the older end of the age range 100-500 Myr. Evolution model calculations of \citet{Leggett2008} predict a mass of $28\,M_{\rm Jup}$ for an age of 500\,Myr.

\section{Results}

We obtained high-quality G141 spectral time series for \hnb. Our spectra cover the wavelength range from 1.1 to $1.7\micron$ including the $J$ photometric band and most of the the $H$ band, as well as the water and methane absorption bands. The spectra of \hnb{} were dominated by water absorption near 1.1 and $1.4\micron$, which is consistent with an L/T transitional spectral type. The SNRs of the spectral are $\sim170$ at the bright $J$ and $H$ band peaks and $\sim40$ at the faint $1.4\micron$ water absorption bands.

\begin{figure}[t!]
  \centering
  \plotone{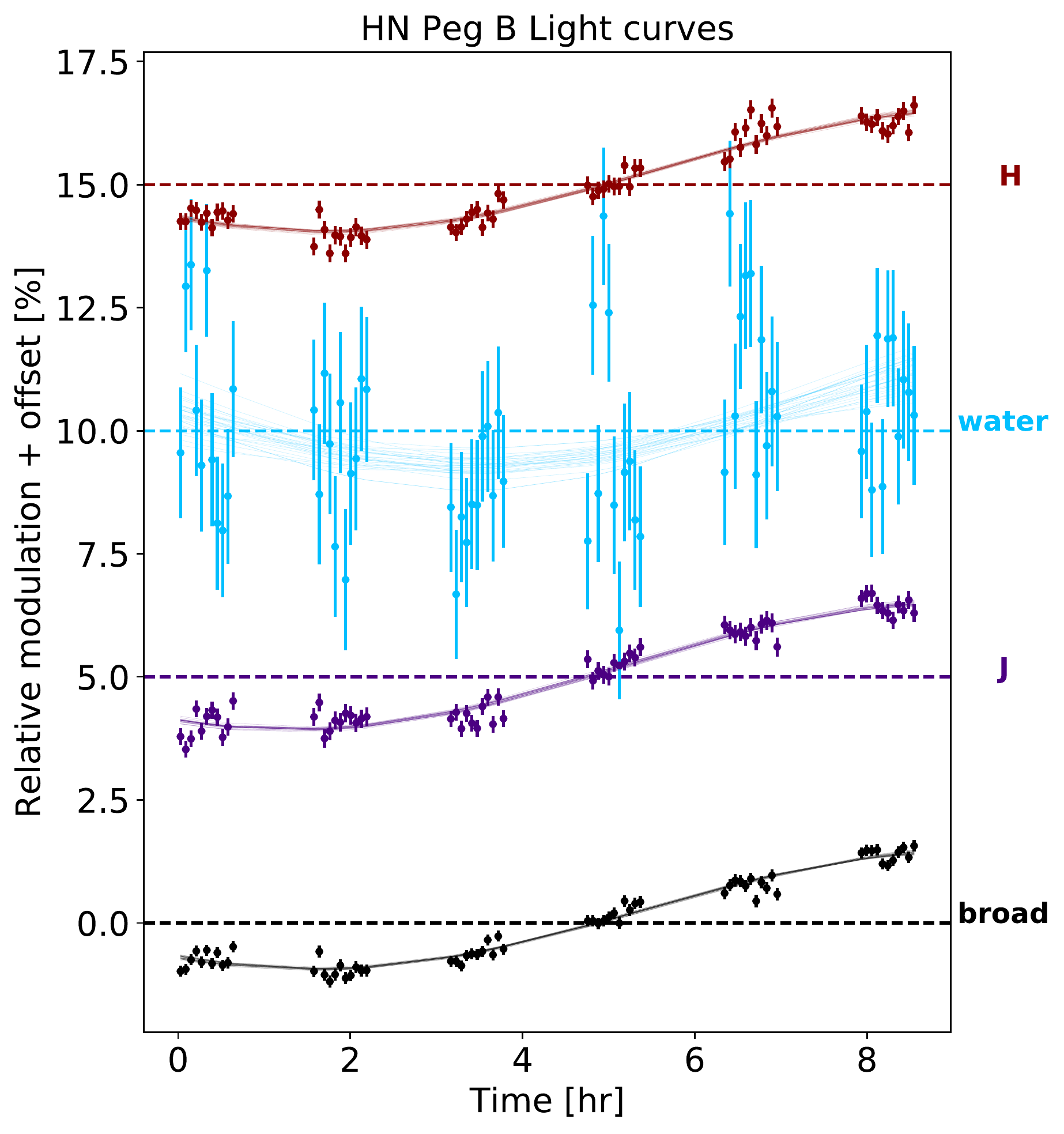}
  \caption{HN Peg B's light curves in WFC3 G141 band pass. Observed $1.1$--$1.7$ \micron{} broad, $J$, water, and $H$ bands light curves are plotted in dots with errorbars. For each band, we also plot 30 sinusoids that are randomly selected from the posterior distributions of the sinusoidal fits (\S\ref{sec:sine}).  }
  \label{fig:lightcurves}
\end{figure}

We constructed near-infrared light curves of \hnb\, for
1.1-$1.7\micron$ white light as well as $J$, $H$\footnote{ The
  transmittance of G141 grism falls short on the red end comparing to
  that of MKO $H$ filter so our $H$ band photometry does not include
  flux above 1.7 \micron.} and the water absorption band
(Figure\,\ref{fig:lightcurves}). $J$ and $H$ photometric points were
integrations of the product of the spectrum and MKO $J$/$H$ filter
transmittances. The water band photometry was integrated from 1.37
\micron{} to 1.41 \micron{}.  We achieved an SNR of $\sim 850$ for the
1.1-1.7 \micron{} broadband light curve. \hnb's light curves showed a
rising trend in amplitude over the 8.6 hours time span with the lowest
point at the second
orbit.

\subsection{Sinusoidal fit}
\label{sec:sine}
The main purpose of our investigation is not the detailed analysis of the light curve morphology but the study of the wavelength-dependence of the variations. Nevertheless, in order to explore a likely range of periodicity in the observed variations, we fit a simple sinusoidal function to the light curves of \hnb. This is motivated by \textit{Spitzer} photometric studies \citep[][]{Metchev2015,Apai2017}. In the sinusoidal fit, we used \textit{Spitzer} light curve period estimation $18\pm4$ hr as a prior. We found the best-fit sinusoid's amplitude to be $1.206\%\pm0.025$\% and its period to be $15.4\pm0.5$ hours. This near-infrared amplitude is greater than the \textit{Spitzer} 3.6 \micron{} band amplitude \citep[$0.77\%\pm0.15$\%][]{Metchev2015} and very similar to the 4.5 \micron{} band's amplitude \citep[$1.1\%\pm0.5$\%]{Metchev2015}. However, we note that light curves are expected to significantly evolve between the two observations, and WFC3/G141 and \textit{Spitzer} probe different atmosphere altitudes \citep{Yang2016}. Our best-fit period agrees within the uncertainties with the period of $18\pm4$ hours estimated from the 21h-long \textit{Spitzer} light curve presented in \citet{Metchev2015}.  We note that our observation covers about 50\%-60\% of a full rotation, which may bias the rotation period estimation. In fact, \citet{Vos2017} used full phase covered \textit{Spitzer} light curve and found longer period for brown dwarf WISEJ0047 than that measured from partially phase covered HST light curve \citep{Lew2016}. The uncertainty of the period can be under-estimated, because we only propagated photometric uncertainty, but did not account for the possibility that light curve deviated from a single sinusoid.

In order to explore possible wavelength dependent phase shifts (e.g., \citealt[][]{Buenzli2012,Apai2013,Yang2016}) we performed similar sinusoidal fits to each of the three spectral bands shown in Figure~\ref{fig:lightcurves}. In these fits, we kept the periods fixed to the best-fit value from the broadband results (15.4~hr) and investigated the possible phase and amplitude differences between the bands. For the $J$ and $H$ bands, we derived amplitudes of $1.28\%\pm0.03\%$ and $1.25\%\pm0.03\%$ and phases of $0.663\pm0.005$ and $0.639\pm0.005$, respectively. Therefore, the sinusoidal amplitudes of $J$ and $H$ bands are statistically identical within $1\sigma$ and are slightly larger than the broadband value. $J$ and $H$ bands have a phase difference of 2.4\% ($3.4\sigma$). However, we consider this possible phase difference tentative given the very basic light curve modeling applied here, which also affects our uncertainty estimates. In contrast to the $J/H$ bands, the $1.4\micron$ water band light curve does not have enough precision to produce a reliable phase measurement (see Figure~\ref{fig:lightcurves}).

\begin{figure}[t!]
  \centering
  \plotone{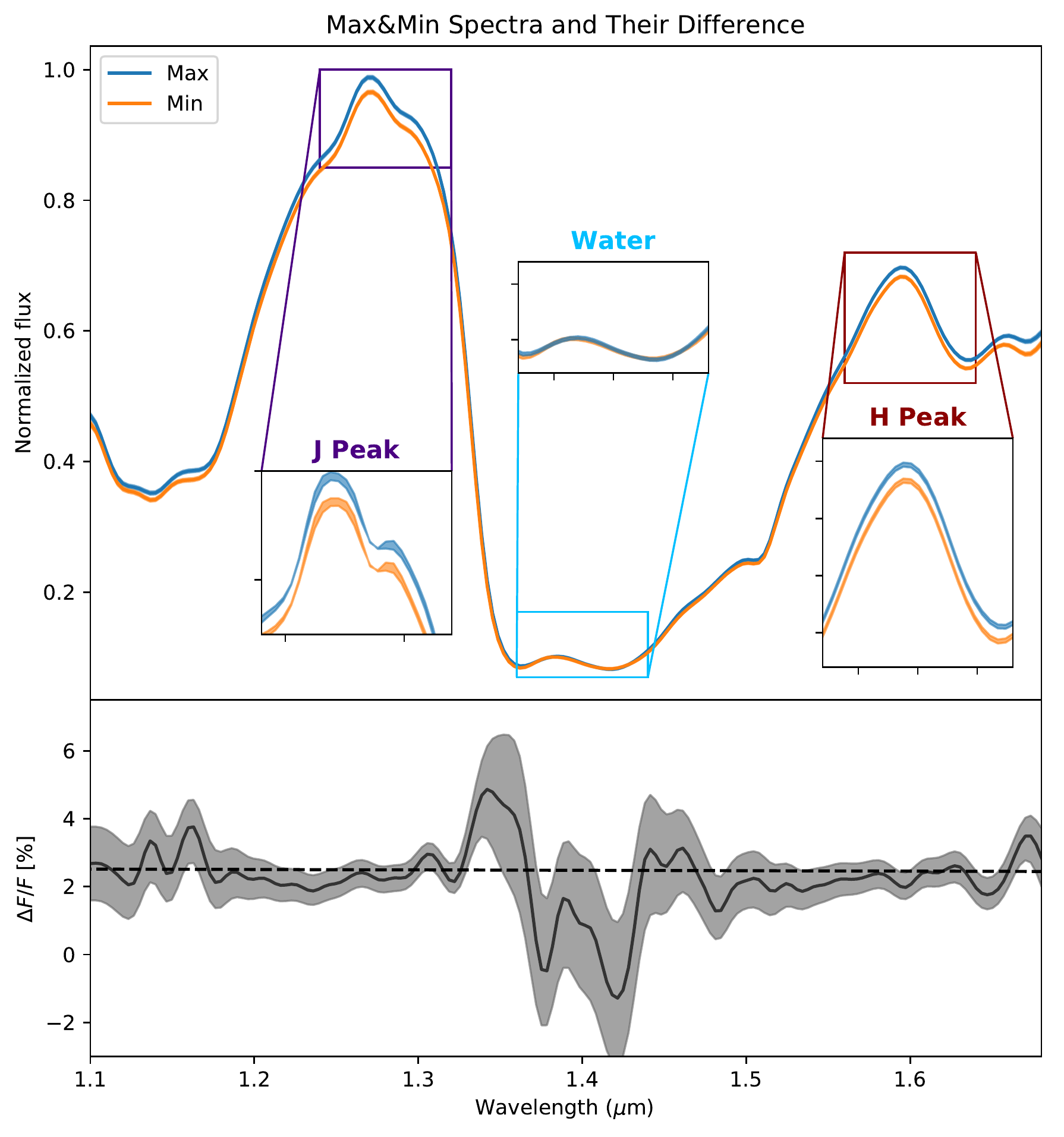}
  \caption{Wavelength dependence of HN Peg B's rotational modulation amplitudes. We median combined the 8 brightest (blue) and 8 faintest (orange) spectra, and calculated their difference based on Equation \ref{eq:2}. Three sub-panels highlight $J$, water, and $H$ bands. Spectra and their difference are all smoothed using a Gaussian kernel with $\sigma=1.5$ pixels.} \label{fig:diff}
\end{figure}

\subsection{Spectral modulations}

The high SNR spectral time series allowed us to investigate the wavelength dependence of the rotational modulations. Following a method introduced by \citet[][]{Apai2013} we selected the eight spectra closest to the brightest segment and the eight spectra closest to the faintest segment of the light curve and median-combined each sets. In Figure~\ref{fig:diff}, we plot the two median spectra and their relative difference. Following \citet{Buenzli2015}, we defined spectral modulation as $\dfrac{\Delta F_{\lambda}}{F_\lambda}$
\begin{equation}
  \label{eq:2}
  \frac{\Delta F_\lambda}{F_\lambda} = \frac{F_{\lambda, \mathrm{max}}
  - F_{\lambda, \mathrm{min}}}{F_{\lambda, \mathrm{mean}}}
\end{equation}
and plot it in the bottom panel of Figure~\ref{fig:diff}.  Two key results are immediately apparent from Figure\,\ref{fig:diff}. First, the spectral modulations are {\em very gray}, essentially identical in the $J$- and $H$-bands. Second, the amplitude is significantly reduced in the $1.4\micron$ water absorption band: the relative water band maximum-to-minimum difference (wavelengths ranging from 1.37 to $1.41 \micron$) is only $0.80\pm0.41\%$ while this difference outside the water absorption band is $2.56 \pm 0.06\%$. The difference in the water band ($1.37\micron < \lambda < 1.41\micron$) is $4.36\,\sigma$ below that outside of water band and only $1.9\,\sigma$ level above zero. Similar reductions of modulation amplitude in water absorption band have been previously found in all three L/T transition brown dwarfs with HST/G141 time-resolved spectroscopy \citep{Apai2013, Yang2014, Buenzli2015}.

\section{Discussion}

\begin{figure}[t]
  \plotone{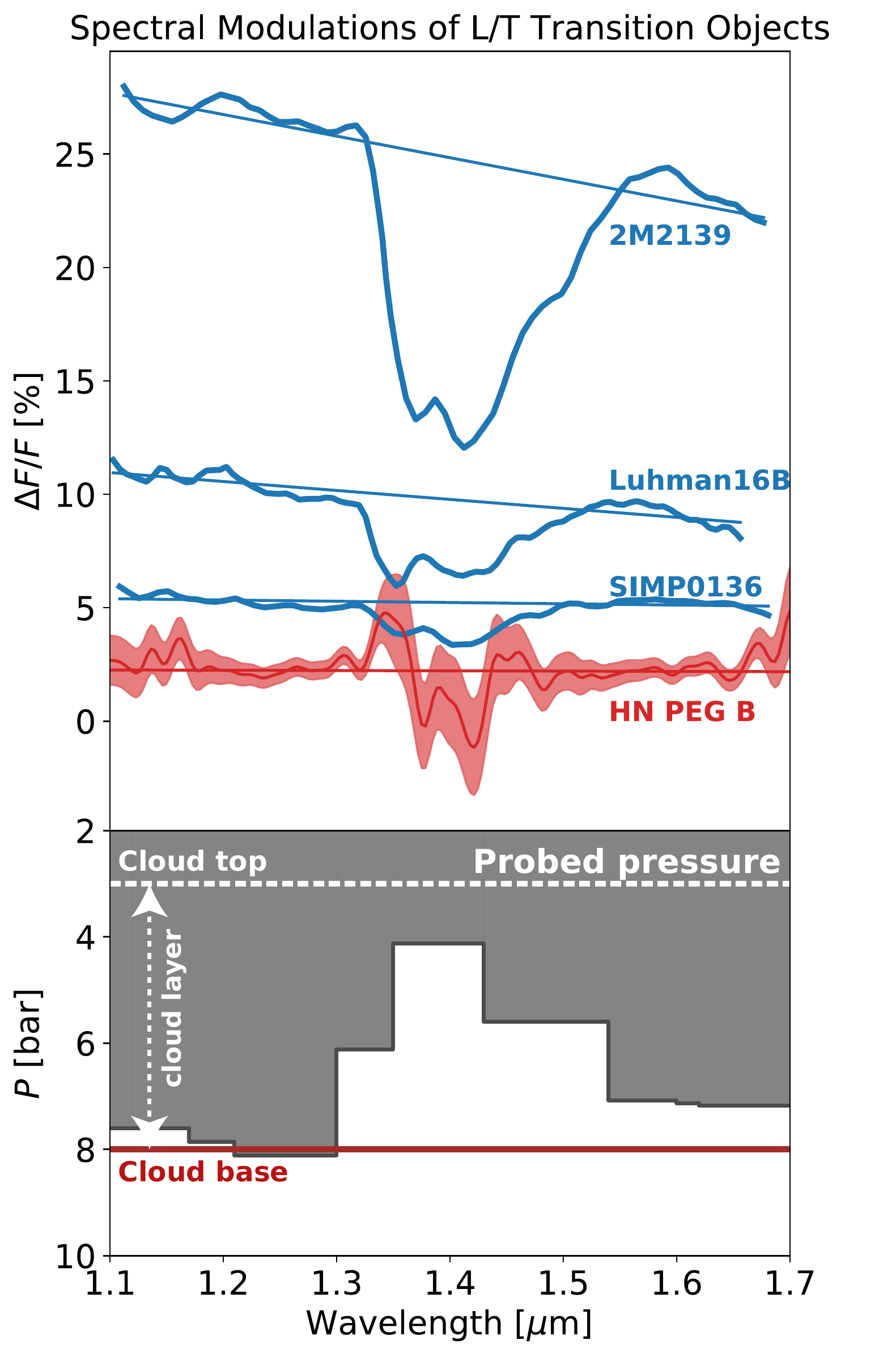}
  \caption{Comparison of HN\,Peg\,B with L/T transition brown dwarfs that have WFC3 G141 time-resolved observations. \emph{Upper:} Spectral modulations defined in equation \ref{eq:2} are plotted. Lines are best-fit linear trends for spectral modulations out of 1.4\,\micron\, water absorption band. \emph{Lower:} Vertical spans of the atmosphere that are probed by different wavelength light curves are indicated in gray. (Pressure levels are calculated specifically for SIMP0136. Wavelength ranges that were not listed in \citet{Yang2016} are interpolated with adjacent bands. Red solid and dashed line show cloud base and top, respectively.)}
  \label{fig:compare}
\end{figure}

\subsection{Spectral slopes and Amplitudes}
Our WFC3/G141 time-resolved spectroscopy of \hnb{} enlarges the small sample (size of 3) of L/T transition brown dwarfs that have precise spectro-photometric rotational modulation measurements. In addition, HN\,Peg\,B is different from the existing sample of field L/T transition brown dwarfs because it is a {\em companion to a star}, with a star-to-companion mass ratio estimated to be $>30$. Here, we compare HN\,Peg\,B with L/T transition brown dwarfs 2M2139, SIMP0136 \citep{Apai2013, Yang2014}, and Luhman16\,B \citep{Buenzli2015} to explore the wavelength dependence of their rotational modulations.  We compiled the four objects' spectral modulation curves in Figure~\ref{fig:compare}. The spectral modulation curves of 2M2139, SIMP0136, and Luhman 16 B are reproduced from literature. The four objects' broadband modulations amplitudes range from $2.56\%$ to $24.9\%$, however, their modulations all follow a similar pattern. Each of the four rotational amplitude curves show a significant decrease in the 1.4 \micron{} water absorption band. We identify three quantities to characterize spectral modulations: the 1.1--1.7 \micron{} broadband modulation amplitude, the difference in the 1.4 \micron{} water absorption with respect to the adjacent continuum bands $\dfrac{\Delta F_{\mathrm{out}}}{F_\mathrm{out}} - \dfrac{\Delta F_{\mathrm{in}}}{F_\mathrm{in}}$, and the $\dfrac{\Delta F_{\lambda}}{F_\lambda}$ modulations' spectral slope measured outside of the water absorption band. We follow \citet{Yang2014} to define $\dfrac{\Delta F_{\mathrm{in}}}{F_\mathrm{in}}$ as the weighted average $\dfrac{\Delta F_{\lambda}}{F_\lambda}$ with HST/WFC3 filter F139M's throughput and $\dfrac{\Delta F_{\mathrm{out}}}{F_\mathrm{out}}$ with that of filter F127M and F153M \citep{Dressel2017}. The results are listed in Table~\ref{tab:compare}.

\begin{deluxetable*}{lcccccc}
  \tablenum{1}
  \tablecaption{Spectral modulation characteristics in the four L/T
    transition brown dwarfs with HST/G141 Time-resolved
    Spectroscopy.\label{tab:compare}}
  \tablehead{
    \colhead{Object}&
    \colhead{Spec. Type}&
    \colhead{$J-H$}&
  \colhead{Broadband Amplitude} &
  \colhead{Slope} &
  \colhead{$\dfrac{\Delta F_{\mathrm{out}}}{F_\mathrm{out}} -
    \dfrac{\Delta F_{\mathrm{in}}}{F_\mathrm{in}}$}&
  \colhead{References}\\
  \colhead{} &
  \colhead{} &
  \colhead{mag}&
  \colhead{(\%)} &
  \colhead{(\%/\micron{})} &
  \colhead{(\%)}&
  \colhead{}
  }
\startdata
HN Peg B\tablenotemark{a}  & T2.5 & 0.46 & $2.56\pm0.06$   &
$-0.13\pm0.28$  & $1.7\pm0.4$  & 1, 2 \\
2M2139    & L8.5+T3.5 & 0.55 & $24.91\pm0.02$  & $-9.52\pm0.13$   &
$11.6\pm0.1$  & 3, 4\\
SIMP0136  & T2.5 & 0.69 &$5.23\pm0.03$   & $-0.55\pm0.13$  &
$1.6\pm0.2$   & 4, 5\\
Luhman16B & T0.5 & 0.83 & $9.9\pm0.03$    & $-3.94\pm0.14$  & $3.2\pm0.2$   & 6\\
\enddata
\tablenotetext{a}{HN Peg b's modulation amplitude should be considered
  a minimum since our observations covered only $\sim$50-60\% of the rotational phase space.}
\tablerefs{(1) \citet{Luhman2007}; (2) \citet{Leggett2008}; (3)
  \citet{Burgasser2010}; (4) \citet{Cutri2003}; (5)
  \citet{Faherty2009}; (6) \citet{Burgasser2013}}
\end{deluxetable*}

Among the four brown dwarfs, HN Peg B has the lowest modulation amplitude as well as the flattest spectral modulation slope. HN Peg B's measured spectral modulation slope -- outside of $1.4\micron$ molecular absorption bands -- agrees with wavelength-{\em independent} modulations.  The broadband amplitudes and spectral modulation-wavelength slopes of L/T transition brown dwarfs follow a correlation that was initially recognized by \citet{Lew2016} using a sample of six brown dwarfs that have WFC3/G141 rotational modulation measurements.  The low amplitude and flat spectral modulation spectral slope of HN\,Peg\,B further reinforces this empirical trend.

\subsection{Cloud top heights}

The lower modulation amplitude in the 1.4\micron{} water absorption band can be explained by a model proposed by \citet[][sample including L and L/T transition brown dwarfs]{Yang2014}, who used an analytical approximation as well as a radiative transfer model to demonstrate the mechanism.  In this model, the relative modulation amplitude in and out of the water absorption band is approximately given by $\epsilon = \exp(\tau_{\mathrm{cont.}} - \tau_{\mathrm{water}})$, where $\tau_{\mathrm{cont.}}$ and $\tau_{\mathrm{water}}$ are the optical depths measured from the top of atmosphere to the cloud layer that introduces the modulations.  We explored the implications of this model in the context of the spectral slopes observed in Table 1.  Along with their spectra in Figure~\ref{fig:compare}, we also plot the pressure levels the different wavelengths probe the atmospheres --- 80\% of the flux emerge from the gray shaded region and above in lower panel of Figure~\ref{fig:compare}.  We adopt the same pressure level results for SIMP0136 from \citet{Yang2016}, because of the similarities of spectral type between \hnbsp and SIMP0136.
For example, in the $J$-band continuum window, observations will typically probe down to about 7-9 bar pressures, while in the 1.4\micron{} water band, observations probe only down to about 2-4 bar pressure levels. Therefore, the effective top of atmosphere near 1.4\micron{} is $\sim 3$ bar lower in pressure than it is in the continuum band. This in turn implies that $\tau_{\mathrm{cont.}} - \tau_{\mathrm{water}}$ is negative and modulation in water absorption band is smaller than it is in the continuum band. The relative modulations in and out of the 1.4 \micron{} water absorption band for \hnbsp agree well (within $1\sigma$ uncertainty) with those reported for the other three L/T transition brown dwarfs. This similarity suggests that the vertical cloud structure for HN\,Peg\,B is not different from the three L/T transition brown dwarfs studied previously. These four brown dwarfs demonstrate that, in spite of some differences, most L/T brown dwarfs will likely share very similar vertical cloud structures. Our small sample only spans a limited range of surface gravities (mid to high) and does not yet allow direct comparisons to model predictions. In the near future, with the completion of the Cloud Atlas program, a four times larger sample that — will also include several low-gravity brown dwarfs — will enable breaking the degeneracies between surface gravity and other atmospheric parameters.

\subsection{Constraints on Cloud Particle Sizes}
We also investigate the characteristic particle size in the condensate clouds assuming that Mie scattering extinction is the primary source of modulations, following \citet[][]{Hiranaka2016}. While the best constraints on atmospheric aerosol particle sizes come from broad wavelength coverage, the flat slope over our WFC3 spectral range does allow us to place a constraint on the minimum particle size of the clouds. We used the same model as \citet{Schlawin2017} in which clouds that introduce the modulations are made of spherical forsterite grains and optically thin, and the dust particle size is described by a log-normal distribution characterized by the median grain radius $r$ and the scale parameter $\sigma_s$. In this model, the spectral modulation amplitudes linearly scale with the Mie extinction coefficients.  We find that in order to reproduce a flat modulation spectral slope, the model requires relatively large characteristic particle sizes ($r > 1.0 \micron$). The particle size we find for HN\,Peg\,B is significantly greater than those for dusty late-L-type brown dwarfs \citep[$\sim0.2-0.4\micron$,][]{Lew2016,Schlawin2017}, but similar to some less-varying L dwarfs (2M1507, \citealt{Yang2014}, LP261-B, \citealt{Manjavacas2017}).

\subsection{Toward High-Contrast Time-resolved Spectroscopy}
In addition to the astrophysical results, our study illustrates the complicating factors (contamination and complex spectrally dispersed point spread function) introduced by nearby bright companions in WFC3 G141 observations. Importantly, even though the bright source HN\,Peg\,A was more than $40''$ away, well outside the field-of-view, these effects, without mitigation in post-processing as detailed in \S\ref{sec:removal}, would have seriously impacted our observations. To remediate, we demonstrate that empirical interpolation can effectively correct for such systematics and allows nearly photon-noise limited precision. We describe a spatially periodic light pattern on the detector and present two lines of evidence supporting that it originates from a spectrally dispersed Fraunhofer diffraction pattern from a circular aperture. First, the traces of the stripes converge to the same point (outside our images). Second, we show that the median brightness of the stripes decay as $1/N^3$, where $N$ is the stripe number. Therefore, an optical contamination correction model will likely improve WFC3/G141 time-resolved observations for which the targets are close binaries \citep[e.g.,][]{Buenzli2015, Beatty2017} or have nearby bright sources. Future versions of WFC3/G141 PSF simulators (e.g., aXeSIM, \citealt{Kummel2009}; Wayne, \citealt{Varley2017}) may also take this effect into account.

\section{Summary} 
The key results of our study are as follows:

\textbf{(i)} Our study demonstrates an empirical image processing method that can remove the contamination from the spectrally dispersed side lobes of the point-spread functions of bright, close companions to the target in G141 observations.

\textbf{(ii)} We present time-resolved spectroscopy of the rotational modulations in the L/T transition brown dwarf mass companion HN\,Peg\,B.  We confirm the presence of rotational modulations in the light curve of \hnb. The 1.1\micron{}-1.7\micron{} broadband modulation amplitude and period are $1.206\pm0.025\%$ and $15.4\pm 0.5$\,hours, respectively, which are consistent with those reported in the \textit{Spitzer Space Telescope} photometric study by \citet[][]{Metchev2015}, considering the different wavelengths of observation.  The temporal baseline of our observations corresponds to about 60\% of the estimated rotational period of \hnb. Our light curve appears to probe the minimum of the light curve but not its peak.

\textbf{(iii)} We found that the rotational amplitude outside the water absorption band is nearly independent of wavelength: our measurements are consistent with an amplitude of $1.25\%$ in the $\lambda=1.1-1.65 \mu$m wavelength range, with the exception of the $1.4\mu$m water band. We found a significantly reduced variability amplitude in the water band. The modulation amplitude in the water absorption band is lower than that of the continuum by more than $4\sigma$ level.

\textbf{(iv)} \hnbsp is a non-tidally locked brown dwarf companion for which rotational spectral mapping is possible. We found that its rotational modulations resemble closely those observed in higher-gravity and (presumably) older T2 brown dwarfs (\object{2M2139}: \citealt[][]{Apai2013}; \object{Luhman 16}: \citealt[][]{Buenzli2015}) and that of the likely planetary-mass but unbound brown dwarf SIMP0136 \citep[][]{Apai2013}.

\textbf{(v)} Assuming Mie scattering, the fact that \hnb's rotational modulation amplitude is very close to wavelength-independent argues for the characteristic dust particle size of the cloud to exceed 1\micron{}.

Our study increased the small sample of L/T transition brown dwarfs with time-resolved spectroscopy. We show that the nature of the rotational modulations in the first L/T companion to a star resembles those observed in field brown dwarfs, suggesting that cloud vertical structures are similar for companions to stars and for individual brown dwarfs.

\acknowledgments We acknowledge the anonymous referee for a constructive report that improves the manuscript. We acknowledge Dr. Sandy Leggett for valuable discussion on HN Peg B's spectral fitting and Dr. Peter McCullough for helpful discussion on the cause of the image contamination. Y.Z. acknowledges support in part by the NASA Earth and Space Science Fellowship Program - Grant “NNX16AP54H”. D.A. acknowledges support by NASA under agreement No. NNX15AD94G for the program Earths in Other Solar Systems. Support for Program number 14241 was provided by NASA through a grant from the Space Telescope Science Institute, which is operated by the Association of Universities for Research in Astronomy, Incorporated, under NASA contract NAS5-26555. Based on observations made with the NASA/ESA Hubble Space Telescope, obtained in GO program 14241 at the Space Telescope Science Institute.

\software{aXe, aXeSIM \citep{Kummel2009}, IRAF \citep{Tody1986, Tody1993}, REDSPEC pipeline \citep{Prato2002}, Numpy\&Scipy \citep{VanderWalt2011}, Matplotlib \citep{Hunter2007}, IPython \citep{Perez2007}, Astropy \citep{Robitaille2013}, scikit-image \citep{scikit-image}}

\bibliographystyle{yahapj}

\end{document}